\documentstyle[12pt]{article}
\begin{document}

\hfill{FTUV/93-54}

\begin{center}
{\LARGE{\bf Representations of q-Minkowski\\
space algebra}}
\end{center}

\vspace{1cm}

\begin{center}
{\large{\bf P.P. Kulish \footnote{On leave of absence from St.Petersburg
Branch of the Steklov Mathematical Institute of the Russian Academy of
Sciences.}}}
\end{center}

\begin{center}
{\bf \it Departamento de F\'{\i}sica Te\'orica and IFIC,\\
Centro Mixto Universidad de Valencia-CSIC\\
46100 Burjassot (Valencia), Spain}
\end{center}

\rightline{{\it To Ludwig Dmitrievich Faddeev}}
\rightline{{\it on his sixtieth anniversary}}

\vspace{2cm}

\begin{abstract}
The properties of the quantum Minkowski space algebra are discussed. Its
irreducible representations with highest weight vectors are constructed and
relations to other quantum algebras: $su_{q}(2)$, $q$-oscillator, $q$-sphere
are
pointed out.
\end{abstract}

\section{Introduction}

The intensive development of quantum integrable systems in two dimensions
in the last fifteen years was strongly influenced by the discovery of the
quantum inverse scattering method (QISM) \cite{1}.
In this method the important concepts and names were introduced which are
rather well known these days: the $R$-matrix, Yang-Baxter equation (YBE),
quantum determinant, different forms of Bethe Ansatz (algebraic, analytic,
functional) etc. (for reviews see \cite{2}). One of the mathematical structures
extracted from the QISM was the notion of quantum group introduced in
\cite{3,4}. Although the development of the QISM is going continuously with
impressive progress in the evaluation of correlation functions we see much
higher activity in the quantum group field. It is certainly related to the
fact that many properties of quantum groups (quasitriangular Hopf algebras) are
very similar to those of Lie groups and Lie algebras. However, it seems that
another reason is the formulation of the quantum group theory in terms of the
$R$-matrix formalism \cite{5}.

The quantum group theory is used to describe more elaborated symmetries of
physical models. It also gives rise to many explicit examples of
non-commutative geometry (see e.g. \cite{6}). For theoretical physicists it is
very interesting to use the amazing possibility of constructing a
non-commutative space-time according to the well-defined
group-theoretical scheme where the Minkowski space $M$
appears as a factor space of the Poincar\'e group $P$ over
the Lorentz group $L$, $M$ $\sim$ $P/L$. One of the difficulties for
physicists who
start working in this field is that the corresponding manifolds are absent in
quantum groups (QG).
Instead of them one has to use non-commutative analogs of algebra of functions
on group ${\cal F}(G)$.

One of the definitions of quantum (or $q$-deformed) Lorentz group ${\cal
L}_{q}$
and corresponding
$q$-Minkowski space algebra ${\cal M}_{q}$ was given in \cite{7}. According to
this definition ${\cal M}_{q}$ is an associative algebra (with
$\ast$-operation)
generated by four elements: $\alpha , \beta , \gamma ,\delta$ which satisfy
the relations ($\lambda =q-q^{-1}$)

\begin{equation}
\begin{array}{ll}
\alpha \gamma = q^{2} \gamma \alpha, \quad & \quad [\beta , \delta ]=
-(\lambda /q) \alpha \beta ,\\

\alpha \beta = q^{-2} \beta \alpha , \quad & \quad [\gamma ,\delta
]=(\lambda/q)
\gamma \alpha ,\\

\alpha \delta = \delta \alpha, \quad & \quad [\beta , \gamma ]=(\lambda/q)
(\alpha \delta - \alpha^{2}).
\end{array}
\end{equation}

\noindent
The transformation of ${\cal M}_{q}$ under (co)action of the $q$-Lorentz group
${\cal L}_{q}$ will be defined in the next Section as well as ${\cal L}_{q}$
itself.

The aim of this paper is to construct irreducible representations of ${\cal M}
_{q}$ with a highest weight vector (vacuum) and to discuss some algebraic
problems related to non-commutative differential geometry (NCDG)
of ${\cal M}_{q}$.

\section{Reflection equation definition of q-Minkowski space algebras}

The starting point of the papers \cite{7} is the relation of the Lorentz
group $L$ to $SL (2, C)$ and the spinorial construction of the Minkowski space
coordinates $x^{\mu}$. It is also possible to use for the definition of a
$q$-Minkowski space the well-known $2 \times 2$ matrix relation expressing the
$2 \longrightarrow 1$ homomorphism between $SL(2, C)$ and the Lorentz group,

$$
x^{\mu} \longrightarrow x^{'\mu} = \Lambda^{\mu}_{\nu} x^{\nu}, \quad
\sigma_{\mu}
x^{\mu} \longrightarrow \sigma_{\mu} x^{'\mu} = A \sigma_{\mu} x^{\mu}
A^{\dagger},
$$

\begin{equation}
A, A^{\dagger} \in SL(2, C)
\end{equation}

\noindent
in the framework of the $R$-matrix formalism \cite{5} and reflection equations.
The covariance properties of $\sigma_{\mu} x^{\mu}$ expressed by (2)
are translated in this way to the deformed case. In this manner, all relations
which define the quantum De Rham complex of ${\cal M}_{q}$ (coordinates (1),
$q$-derivatives, $q$-1-forms) proposed in \cite{8} can be written in compact
unified
form \cite{9}.

In order to $q$-deform the transformation (2) it is natural to consider instead
of $\sigma_{\mu} x^{\mu}$ just a $2 \times 2$ matrix $K$, the entries of
which are the generators of the $q$-Minkowski space algebra ${\cal M}_{q}$ in
question. Following \cite{7} we introduce two isomorphic but mutually
non-commuting copies of the quantum group $SL_{q} (2, C)$. The commutation
relations among generators of these quantum groups $(a, b, c, d;\tilde{a},
\tilde{b}, \tilde{c}, \tilde{d})$ in matrix form \cite{5,7} look like this:

\begin{equation}
R_{12} M_{1} M_{2} = M_{2} M_{1} R_{12},
\end{equation}

\begin{equation}
R_{12} \tilde{M}_{1} \tilde{M}_{2} = \tilde{M}_{2} \tilde{M}_{1} R_{12},
\end{equation}

\begin{equation}
R_{12} M_{1} \tilde{M}_{2} = \tilde{M}_{2} M_{1} R_{12},
\end{equation}

\noindent
where

\begin{equation}
M=
\left(
\begin{array}{ll}
a \; b\\
c \; d\\
\end{array}
\right)
,\, det_{q} M=ad-qbc=1,\, q \in R
\end{equation}

\noindent
and $\tilde{M}$ is used for an isomorphic copy of $SL_{q}(2, C)$, so $det_{q}
\tilde{M} = \tilde{a} \tilde{d} - q \tilde{b} \tilde{c} =1$. We use the
standard notations for the QISM as well as for the YBE and the quantum group
theory (cf. \cite{5,6}) e.g. $M_{1}=M \otimes I, M_{2}= I \otimes M$ and
$R_{12}$ are $4 \times 4$ matrices in $C^{2} \otimes C^{2}$, the $R$-matrix is
well-known for the $SL_{q}(2, C)$ \cite{5}. This set of generators $(a, b,...)$
define the quantum Lorentz group ${\cal L}_{q}$. The $\ast$-operation or
reality condition for ${\cal L}_{q}$ is $M^{\dagger}=\tilde{M}^{-1}$ \cite{7}.

The transformation of the generators $(\alpha , \beta , \gamma , \delta )$ of
${\cal M}_{q}$ is written as in the classical case (2)

\begin{equation}
\phi ; K =
\left(
\begin{array}{ll}
\alpha \; \beta\\
\gamma \; \delta
\end{array}
\right)
\longrightarrow
K'= M K \tilde{M}^{-1}
\end{equation}

\noindent
where it is assumed that the entries of $K$ commute with those of $M$ and
$\tilde{M}$. This map $\phi$ is a coaction of the Hopf algebra ${\cal L}_{q}$
on the algebra ${\cal M}_{q}$ and as such it must be a homomorphism. To see
that the
defining relations (1) of ${\cal M}_{q}$ are preserved by map (7) we can
write them
in the form of appropriate reflection equation (RE) (cf. \cite{10,11} and
refs. therein)

\begin{equation}
R_{12} K_{1} R_{21} K_{2} = K_{2} R_{12} K_{1} R_{21}
\end{equation}

\noindent
where

\begin{equation}
R_{21}= {\cal P} R_{12} {\cal P}
\end{equation}

\noindent
and ${\cal P}$ is the permutation operator in $C^{2} \otimes C^{2}$. Now it is
just matrix algebra exercise to check that (8) is invariant under (7) taking
into account the defining relations (3)-(5) of ${\cal L}_{q}$ (it is so for
any $GL_{q} (n)$ or quantum group, defined by the $R$-matrix relations).

Using the reality condition $\tilde{M}^{-1}=M^{\dagger}$ of ${\cal L}_{q}$, the
coaction (7) may be written as (2) $K'=MKM^{\dagger}$. This means that the
reality
($\ast$-operation) of the $q$-Minkowski space ${\cal M}_{q}$ may be expressed
as in the classical case (2) by the hermiticity of $K$ i.e.
$K^{\dagger}=K$. This requirement is
consistent with (7) and the RE (2), because the latter one goes into itself
after hermitian conjugation due to the $R$-matrix properties $R_{12}^{\dagger}=
R_{21}.$

The centrality of the following two elements of ${\cal M}_{q}$

\begin{equation}
c_{1} \equiv q^{-1} \alpha + q \delta ,
\end{equation}

\begin{equation}
c_{2}= \alpha \delta - q^{2} \gamma \beta ,
\end{equation}

\noindent
easily follows from the invariance property of the $q$-trace \cite{5,6}

\begin{equation}
tr_{q} K= tr D K= q^{-1} \alpha + q \delta , \quad D= diag (q^{-1}, q),
\end{equation}

\noindent
with respect to the quantum group coaction

$$
tr_q K = tr_{q} \{MKM^{-1}\}.
$$

\noindent
By the way, this is true for any $R$-matrix in (3) with $D=tr_{(2)}{\cal
P}_{12}
\left( (R_{12}^{t_{1}})^{-1})^{t_{1}} \right)$ \cite{5,11}.

The algebra defining by the RE (8) appeared also as a lattice current algebra
and as a braid group in \cite{12,13}. Among other
properties of ${\cal M}_{q}$ let us mention an isomorphism of ${\cal M}_{q}$
to ${\cal M}_{1/q}$
and a characteristic equation for $K$ \cite{10,14}

\begin{equation}
q K^{2}-c_{1} K + q^{-1} c_{2}I=0.
\end{equation}

It was shown in \cite{8} that the algebra ${\cal M}_{q}$ can be extended
further according to the non-commutative differential
geometry interpretation by adding algebras of
$q$-forms $\Lambda_{q}$ and $q$-derivatives ${\cal D}_{q}$. Each of them are
generated by four generators as well, which may be put
also in the form of $2 \times 2$ matrices: $dK$
for generators of $\Lambda_{q}$ and $Y$ for ${\cal D}_{q}$. The
additional to (1) set of $4 \times 16$ commutation relations among all these
12 generators \cite{8} can be written in the form of reflection equations
\cite{9}

\begin{equation}
R_{12} K_{1} R_{21} d K_{2}= dK_{2} R_{12} K_{1} R_{12}^{-1},
\end{equation}

\begin{equation}
R_{12} dK_{1} R_{21} dK_{2}=-dK_{2} R_{12}dK_{1} R_{12}^{-1},
\end{equation}

\begin{equation}
R_{12} Y_{1} R_{12}^{-1} Y_{2}= Y_{2} R_{21}^{-1} Y_{1} R_{21},
\end{equation}

\begin{equation}
R_{12} dK_{1} R_{21} Y_{2}= Y_{2} R_{21}^{-1} dK_{1} R_{21},
\end{equation}

\begin{equation}
R_{12} K_{1} R_{12}^{-1} Y_{2} = Y_{2} R_{12} K_{1} R_{21} -q^{2} R_{12}
{\cal P}_{12}.
\end{equation}

The quantum Lorentz group acts on this extended algebra (the quantum De Rham
complex
of ${\cal M}_{q} $) and relations (1), (8), (14)-(18) are invariant
with respect to the transformations

\begin{equation}
K \longrightarrow M K \tilde{M}^{-1}, \; dK \longrightarrow MdK\tilde{M}^{-1},
\; Y \longrightarrow \tilde{M} Y M^{-1} \;.
\end{equation}

\noindent
The corresponding exterior derivative operator can be expressed as the
$q$-trace of matrices of the $q$-$1$-forms and $q$-derivatives

\begin{equation}
d= tr_{q} \{(dK)Y \}.
\end{equation}

It is not easy to comment on several natural questions: is it possible to
extend
the constructed algebra further on, e.g. adding $q$-Grassmannian derivatives of
$\Lambda_{q}$ to have supersymmetric complex? Are there some other
covariant extensions of ${\cal M}_{q}$,
e.g. with the exterior derivative operator (20) which is nilpotent but without
the Leibniz rule \footnote{This possibility for the quantum group
non-commutative
geometry was analyzed by L.D. Faddeev.} as in the case of the quantum
$GL_q(n)$ space? By this reason we shall discuss in
the next section the irreducible representations of ${\cal M}_{q}$.

\section{Representations of ${\cal M}_{q}$}

Due to the fact that one central element (10) is linear in generators one can
change the basis of generators to $\alpha , \beta , \gamma ,$ and $q \tau =
c_{1} = q^{-1} \alpha + q \delta$. Then we will have three non-trivial
commutation relations ($\hat{\lambda} = \lambda /q = (1-q^{-2})$)

$$
\alpha \gamma = q^{2} \gamma \alpha ,\quad \alpha \beta = q^{-2} \beta \alpha ,
$$

\begin{equation}
\beta \gamma = q^{2} \gamma \beta + \hat{\lambda} (l-\alpha^{2})
\end{equation}

\noindent
and centrality of $\tau$ and the quantum Minkowski length $(l \equiv c_{2}$
of (11))

\begin{equation}
l=\alpha \tau - \alpha^{2}/q^{2} - q^{2} \gamma \beta = q^{2} (\alpha \tau -
\alpha^{2} - \beta \gamma )
\end{equation}

\noindent
(second equality follows from (1)).

To analyze the irreducible representations in Hilbert space with positive
metric the simple consequences of (21) are useful ($[n;q] = (q^{n}-1)/(q-1)$)

\begin{equation}
\beta \gamma^{n} = (q^{2} \gamma )^{n} \beta + \hat{\lambda} [n; q^{2}]
\gamma^{n-1} (l-q^{2(n-1)} \alpha^{2}),
\end{equation}

\begin{equation}
\gamma \beta^{n} = (q^{-2} \gamma)^{n} \gamma + \hat{\lambda}/q^{2} [n;q^{-2}]
\beta^{n-1} (l-q^{-2(n-1)} \alpha^{2}).
\end{equation}

\noindent
The mentioned isomorphism of ${\cal M}_{q}$ and ${\cal M}_{1/q}$ is obvious
for (21):

$$
\alpha , l \longrightarrow \alpha', l', \beta \longrightarrow \gamma'/q^{2},
\gamma \longrightarrow \beta'/q^{2},\tau \longrightarrow \tau'/q^{2}, \delta
\longrightarrow (\delta'/q^{2} + \hat{\lambda} \alpha').
$$

We now have an associative algebra with three  generators (one has to remember
the relation (22) of the central element $l$ to $\alpha , \beta , \gamma$ and
$\tau$). Due to the above equivalence ${\cal M}_{q} \sim
{\cal M}_{q^{-1}}$ we can
suppose that $q>1$. The irreducible representations are parametrized by
different values of $l$ and $\tau$.

0. $\alpha = 0$, then other three generators $\beta , \gamma , \delta$
commute among themselves. Hence, $\delta$ is real and arbitrary while $\beta =
\bar{\gamma}$ is arbitrary complex number, $\tau = \delta , l= -q^{2} |
\gamma |^{2}$. This irrep is not faithful. It gives a one-dimensional
representation of the REA (1), (8).

1.  $l- \alpha^{2} =0 \; , \delta=\alpha \: , \beta=\gamma = 0, \tau = \alpha
(1 + q^{-2}).$
This is also a one-dimensional representation, which is not faithful and
corresponds to the stationary point of the coaction (7) of the quantum
'subgroup' $SU_{q}(2)$ of the $q$-Lorentz group ${\cal L}_{q}$.

2.  $l > \alpha^{2}_{0} > 0$, where $\alpha_{0}$ is the vacuum eigenvalue of
$\alpha$ and $\beta |0\rangle=0$. Then from (23) for unnormalized eigenvectors
$|n\rangle=\gamma^{n} |0\rangle$ of $\alpha$ one gets

\begin{equation}
\langle n|n\rangle= (\hat{\lambda})^{n} [n;q^{2}]! \Pi_{k=1}^{n} (l-q^{2(k-1)}
\alpha_{0}^{2}).
\end{equation}

\noindent
For $\alpha_{0} \neq 0$ and $q>1$ this norm will be negative if the integer $n$
is sufficiently big. Because we are looking for irreps in a Hilbert space
with positive metric there must be some $N=D-1$ such that $\| |N+1\rangle \|
\sim
(l-q^{2N} \alpha_{0}^{2})=0$, or

\begin{equation}
\gamma |N\rangle=\gamma |D-1\rangle=0 \; ,
\end{equation}

\noindent
where D is the dimension of the irrep and

\begin{equation}
l=q^{2D} \alpha_{0}^{2}/q^{2} \quad ,\quad \tau=(q^{2D}+1) \alpha_{0}/q^{2}.
\end{equation}

3. $0 < l < \alpha_{0}^{2}$, hence $(l - \alpha_{0}^{2})<0$ and from (23) one
concludes that $\beta$ can not be annihilation operator. So we have to use
(24) supposing that $\gamma |0\rangle=0$. Then for $|n\rangle= \beta^{n}
|0\rangle$ one gets

\begin{equation}
\langle n | n\rangle = (\hat{\lambda}/q^{2})^{n} [n; q^{-2}]!
\Pi^{n}_{k=1}(q^{-2(k-1)}
\alpha_{0}^{2} -l)
\end{equation}

\noindent
where from the same conclusion on finite dimensionality of this irrep follows
like in the previous case

\begin{equation}
l=q^{-2(D-1)}\alpha_{0}^{2}, \quad \tau=(q^{-2D}+1) \alpha_{0}.
\end{equation}

4. $ l \leq 0 \; , \; \alpha \neq 0$, hence $(l-\alpha^{2})<0$ and one has to
use (24) with $\gamma$ as the annihilation operator $\gamma | 0\rangle=0$. Then
one
has for $|n\rangle=\beta^{n} |0\rangle$ just (28)

$$
\langle n | n\rangle=(\hat{\lambda} /q^{2})^{n}[n; q^{-2}]! \Pi_{k=1}^{n}
(q^{-2(k-1)}
\alpha_{0}^{2}-l)
$$

\noindent
which is now positive for any integer $n$. This irrep is infinite dimensional.

Let us comment on obvious relations of the $q$-Minkowski algebra
to the well known $q$-algebras: $sl_{q}(2)$, $q$-oscillator algebra ${\cal
A}(q)$
and $q$-sphere.

\noindent
For $l>0$, considering $\alpha$ as positive element of ${\cal M}_{q}$
and defining generators

\begin{equation}
X_{+}=\alpha^{-1/2} \beta \lambda/(l q^{2})^{1/4} \; , \; X_{-}= \gamma
\alpha^{-1/2} \lambda/(lq^{2})^{1/4},  \; q^{J} = (\alpha/l)^{-1/2}
\end{equation}

\noindent
one gets from (21) the defining relations of the quantum algebra $sl_{q}(2)$

\begin{equation}
q^{J} X_{\pm}= q^{\pm} X_{\pm} q^{J} \quad, \quad [X_{+}, X_{-}]= [2J]_{q}.
\end{equation}

\noindent
This correspondence easily follows from results of \cite{5,12} related $L^{(+)}
,L^{(-)}$ the $2 \times 2$ matrices of quantum algebra $sl_{q}(2)$ generators
with a solution to the reflection equation

$$
K = L^{(-)} (L^{(+)})^{-1}.
$$

\noindent
For $l=0$, one gets the $q$-oscillator algebra ${\cal A}(q_{0})$

\begin{equation}
[A, A^{\dagger}]= q_{0}^{-2N}, [N,A]= -A, [N, A^{\dagger}]=A^{\dagger},
\end{equation}

\noindent
where $q_{0}=q, A^{\dagger} \sim \alpha^{-1/2} \beta,
A \sim \gamma \alpha^{-1/2}, \ \alpha =q^{-2N}$  for $q>1$ and $q_{0}=1/q,
A \sim \alpha^{-1/2} \beta, A^{\dagger} \sim \gamma
\alpha^{-1/2}, \ \alpha=q^{2N}=(1/q)^{-2N}$  for $q<1$.

\noindent
Hence one has always in (32) $q_{0}>1$ and among different inequivalent
irreducible representations of ${\cal A}(q_{0})$ \cite{14,15} in this case only
the
standard oscillator one survives. This is also consistent with the fact that
the central element

\begin{equation}
z=A^{\dagger}A-[N;q_{0}^{-2}]
\end{equation}

\noindent
of the $q$-oscillator algebra is zero for the given realization due to $l=0$.

There is a map of ${\cal L}_{q}$ on the quantum group $SU_{q}(2)$. If we
denote
by $U$ the generator matrix of $SU_{q}(2)$, which satisfies (3) and 'unitarity'
$U^{\dagger}=U^{-1}$, then the mentioned map is: $M \rightarrow U, \tilde{M}
\rightarrow U$. Now both central elements of ${\cal M}_{q}$ are invariant
under the reduced
coaction $K \rightarrow U K U^{\dagger}$. Once $\tau$ and $l$ are fixed the
relations (21), (22) coincide with defining relations of the quantum sphere
algebra \cite{16,17}.

The algebra ${\cal M}_q$ is isomorphic to the $q$-derivative or $q$-momentum
algebra ${\cal D}_q$, hence some of the representations coincide with those
found in \cite{18} for the $q$-deformed Poincare algebra, which has the algebra
${\cal D}_q$ as a subalgebra.

The next step in the representation theory is related to construction
 of a representation in the tensor product of two
irreducible representations. It depends on existence of a bialgebra
( or a Hopf algebra ) structure for the algebra ${\cal M}_q$ or a homomorphism
from ${\cal M}_q$ to ${\cal M}_q \times {\cal M}_q$. Existence of such a map
could be
interpreted physically as the $q$-Lorentz group covariance for two- (or multi-)
particle system. There are few propositions for a possible "coproduct"
$\Delta  : {\cal M}_q \rightarrow  {\cal M}_q \times {\cal M}_q$.
These propositions use:

1. the relation of ${\cal M}_q$ to the quantum algebra $sl_q(2)$, extending it
to isomorphism (modulo some additional requirements ) and introducing the
bialgebra structure through the factorization \cite{12} ( here and below
the indices (1), (2) refer to the factors )

$$
K=L^{(+)}(L^{(-)})^{-1}=L^{(+)}_{(1)}K_{(2)}(L^{(-)}_{(1)})^{-1}.
$$

2. appropriate non-commutativity of the factors in the "tensor product"
${\cal M}_q \times {\cal M}_q$ ("braiding" \cite{13} ), then the matrix
product of two matrices

$$
K=K_{(2)}K_{(1)}
$$

\noindent
will satisfy RE (8) and its entries will generate the algebra isomorphic to
${\cal M}_q$ \cite{11,13};

3. non-commutativity between generators of two factors in such a way, that the
sum of two matrices satisfies (8) \cite{13}

$$
K=K_{(1)}+K_{(2)}.
$$

4. additional matrix ${\cal O}$ constructed from the $q$-Lorentz algebra
generators acting on the first factor such that the matrix

$$
K=K_{(1)}+{\cal O}_{(1)} \times K_{(2)}
$$

\noindent
will satisfy the RE (8) \cite{8}, while the entries of $K_{(1)}$ and $K_{(2)}$
commute (a kind of "undressing" of the preceding case).

Though last two cases look physically reasonable, they together with 2. are
not symmetric with respect to the permutation of factors and not all irreps
of $K_{(1)}$ and $K_{(2)}$ are compatible.

It would be interesting to study which of the above constructed
representations may
be extended to representations of a bigger algebra including ${\cal D}_{q}$ as
well as ${\cal M}_{q}$. This algebra is defined by 8 generators (entries of
$K, Y$) and relations (1) or (8), (16), (18). Introducing explicitly the matrix
elements of $Y$ (we use the notations of \cite{8})

$$
Y=
\left(
\begin{array}{ll}
\partial_{D} & \partial_{A}/q\\
q \partial_{B} & \partial_{C}
\end{array}
\right)
$$

\noindent
one could find, that $\partial_{B}$ and $\partial_{C}$ together with
${\cal M}_{q}$ generate a closed subalgebra. Most of the constructed irreps can
be easily extended to this subalgebra. However, these extensions usually have
singular $q$ dependence for $q \longrightarrow 1$, e.g. for one-dimensional
representation $\alpha=0$ one has $\partial_{C}=0$

$$
\partial_{A}= q^{4}/ \gamma (q^{2}-1),\, \partial_{B}=q^{2}/ \beta (q^{2}-1),\,
\partial_{D}=-q \delta /(q^{2}-1).
$$

In the next Section a central element of this algebra $({\cal M}_{q}, {\cal D}_
{q})$ will be found.

\section{Differential calculus and the $R$-matrix formalism}

The importance of covariant and contravariant vectors (tensors), their
contractions, invariant differential operators etc. is well-known in tensor
calculus and, in particular, in the special relativity theory. Examples of such
elements were given in Sec. 2: if $dK$ is covariant vector, then $Y$ is a
contravariant one, while their contraction, given by the $q$-trace, results in
the invariant operator i.e. the exterior derivative (20). In this Section other
invariant operators and relations among them and the generators of ${\cal M}
_{q}$, ${\cal D}_{q}$ and $\Lambda_{q}$ will be defined in the frame of the
$R$-matrix formalism. Most of the final formulas in terms of components
( generators) can be found in \cite{8}.
However, the $R$-matrix approach demonstrates where one can use a general
$R$-matrix or where such properties as the two eigenvalue characteristic
equation

$$
(\check{R}-q) (\check{R} + 1/q)=0, \quad
\check{R}= q P_{+} -1/q P_{-},
$$

\noindent
or $n=2$, $rankP_{-}=1$ are essential.

First of all let us deduce commutation relations of the linear central
elements of ${\cal M}_{q}$ and ${\cal D}_{q}$ (time variable and corresponding
derivative $\partial_{0}$) with the generators of another algebra. This is
easy,
for it is enough to take the $q$-trace of (18) with respect to the first space
for
$x_{0} \sim c_{1}=tr_{q}K$ and the $q$-trace of slightly transformed (18)
w.r.t. the second space for $\partial_{0}=tr_{q} Y$. One gets

\begin{equation}
Y c_{1}=c_{1} Y + q^{4} I -q^{2} \lambda Y K,
\end{equation}

\begin{equation}
\partial_{0} K = K \partial_{0} + I - q^{-2} \lambda KY,
\end{equation}

\noindent
where the invariance of the $q$-trace and the relations

$$
R_{21}^{-1} = R_{12} - \lambda {\cal P}_{12} = {\cal P}_{12}(q^{-1}I
- [2]_{q}P_{-(12)}),
$$

\begin{equation}
tr_{q(1)} (R_{12} {\cal P}_{12})^{\pm 1} = q^{\pm 2} I_{2}
\end{equation}

\noindent
were used. To get (35) we multiply (18) by $R_{12}^{-1}$ from the left and
by $R_{21}^{-1}$ from the right. The formulas (34), (35) demonstrate that there
is no "naive" reduction $c_1=0$, $\partial_0=0$ to a three dimensional
space algebra.

The algebras ${\cal M}_{q}$ and ${\cal D}_{q}$ are graded ones. It is natural
to introduce a grading operator $N$ with relations

\begin{equation}
[N, K]=K, \quad [N, Y] = -Y.
\end{equation}

On the basis of our experience with the $q$-oscillator algebra ${\cal A}(q)$
(see, e.g. \cite{14}) we know that
once the grading operator $N$ is independent of the generators $A, A^{\dagger}$
it is possible to
find a non-trivial central element (33). The latter one is useful, in
particular,
for the irreducible representation description. The corresponding central
element for ${\cal M}_{q}$ and ${\cal D}_{q}$ has to be related to the
invariant operator (scaling or dilatation)

\begin{equation}
s=tr_{q} K Y.
\end{equation}

\noindent
Its commutativity with $q$-1-forms: $sdK=dKs$ follows easily from the defining
relations (17) and (14) multiplying the first one by $R_{21} K_{2}$ from the
left, by $R_{21}^{-1}$ from the right, using (14) and taking $tr_{q(2)}$.
The relation with coordinates $K$ is more complicated.

Multiplying (18) by $R_{21} K_{2}$ from the left by $R_{21}^{-1}$ from the
right and taking $tr_{q(2)}$, where the index in the brackets like in (36)
refers to the number of space, one gets

\begin{equation}
s K= Ks-q^{2} \lambda KYK+q^{4}K.
\end{equation}

\noindent
For the covariant combination $KYK$ using as an intermediate step

$$
Y_1K_1=q^{-2}tr_{q(2)}(R_{(21)}K_2\check{R}_{(21)}^{-1}Y_2R_{(21)}^{-1})+[2]_qI,
$$

\noindent
we get

\begin{equation}
KYK=q^{-3} Ks + [2]_{q} K + q^{-3} l Y^{\varepsilon},
\end{equation}

\noindent
where the notation for the covariant vector $Y^{\varepsilon}$ is introduced
$(\phi : Y^{\varepsilon} \rightarrow M Y^{\varepsilon} \tilde{M}^{-1})$

\begin{equation}
Y_{1}^{\varepsilon}= [2]_{q}tr_{q(2)} ({\cal P}_{12} P_{-(21)} Y_{2}
R_{21}^{-1})
\end{equation}

\noindent
The resulting relation is

\begin{equation}
s K=q^{-2} K s + K - (1 - q^{-2}) l Y^{\varepsilon},
\end{equation}

\noindent
where $l=c_{2}$ is the $q$-Minkowski length (11), (22).

To cancel 'unwanted' term with $l Y^{\varepsilon}$ in (42) let us calculate the
commutation relation of the $q$-D'Alembertian operator $\Box_{q}$, which
is a central element for ${\cal D}_{q}$,

\begin{equation}
\Box_{q} P_{-(12)}=-q^{-1}P_{-(12)} Y_{1} \check{R}_{12}^{-1} Y_{1} =
-q^{-1}( \partial_{D} \partial_{C} - q^{-2} \partial_{A} \partial_{B} )
\end{equation}

\noindent
with coordinates. This combination is easily going through $K$ once we
construct
such a product of matrices

$$
Y_{3} R_{32}^{-1} Y_{2} R_{13} R_{12} K_{1} R_{21}
$$

\noindent
and apply two times the relation (18) as well as the YBE itself for
different reordering of the $R$-matrices, e.g. $R_{32}^{-1} R_{13} R_{12} =
R_{12} R_{13} R_{32}^{-1}$. Multiplying the final equality by $P_{-(32)}$
from the left and by $R_{21}^{-1} {\cal P}_{23}$ from the right one gets

\begin{equation}
\Box_{q} K=q^{-2} K \Box_{q} - Y^{\varepsilon}.
\end{equation}

\noindent
Hence, the relation of the invariant product $l \Box_{q}$ with $K$ will have
the same 'unwanted' term as in (42). The coefficients $x$ and $y$ in linear
combination $s + x + y l \Box_{q}$ are defined from the requirement that it
commutes with $K$ (and $Y$) as the grading operator $q^{-2N}$ (37): $x=
1/(q^{-2}-1), \, y= (q^{-2}-1)$. As in the $q$-oscillator case the element

\begin{equation}
z= q^{2N}([N; q^{-2}]-s-(q^{-2}-1) l \Box_{q}),
\end{equation}

$$
[n;q]=(q^{n}-1)/(q-1),
$$

\noindent
is central in the algebras ${\cal M}_{q}, {\cal D}_{q}$. It is central also
in the algebra  $\Lambda_{q}$, due to the relations

\begin{equation}
\Box_q (dK)= q^2 (dK) \Box_q , \quad  l(dK) =q^{-2}(dK)l .
\end{equation}

The latter one, for example, follows from (14) multiplying it by
$R_{32}K_{3}R_{23}R_{13}$ from the left, using the YBE, the relation (14) for
$dK_2, K_3$ and finally multiplying by the $q$-antisymmetriser $P_{-(31)}$ from
the left.

Multiplying (18) by $R_{(31)}R_{(32)}K_3R_{(21)}^{-1}{\cal P}_{(13)}$ from the
right
and by $P_{-(13)}$ from the left one gets a "dual" analog of (44)

\begin{equation}
Yl=q^{-2}lY -q^2K_{\varepsilon}
\end{equation}

\noindent
where $K_{\varepsilon}=[2]_qtr_{q(2)}(\check{R}_{(12)}K_1P_{-(12)})$ is the
contravariant vector (the inverse transformation to (41) and proportional to
the inverse of $K$ : $K_{\varepsilon}=lK^{-1}$). This results
to reduction of the $q$-derivatives on the functions $f(l)$
of the invariant length $l$ to a $q$-difference operator
$D_q : f(l)\rightarrow ((1-q^{-2})l)^{-1}(f(l)-f(q^{-2}l))$.
In this manner one could analyze the kernel of $\Box_q$ in the
algebra ${\cal M}_q$. In particular,

$$
trC{\cal P}_{(12)}K_{\varepsilon (2)}R_{(12)}K_1 \in Ker\Box_q, \quad
\hbox{\rm if} \quad trC=0.
$$

\section{Concluding remarks}

The above results demonstrate the rich structure of the $q$-deformed
Minkowski space algebras and usefulness of the $R$-matrix formalism. However,
although the $q$-deformed relativistic one-particle states as unitary
irreducible representations of the quantum Poincar\'e algebra \cite{8} were
defined \cite{18}, the physical features of the future complete theory were
not discussed thoroughly. It is so happened that a $q$-deformed space-time
have been formulated first in frame of the quantum theory without a classical
counterpart. At the same time the exact quantum
relations are often useful for some constructions in the classical theory,
e.g. the quasiclassical limit of the main ingredients of the QISM gave rise
to the classical $r$-matrix and the classical YBE. If we would
introduce Planck's constant just by multiplying the defining relations of
the $q$-algebras by $h$ and then take independent limits $q \rightarrow 1,
h \rightarrow 0$ the resulting relations would be nothing but the
standard Poisson brackets $\{ x_{\mu},p_{\nu} \}= g_{\mu \nu}$ for commuting
coordinates and momenta of the scalar relativistic particle.
If on the other hand the Planck's constant and the
deformation parameter are directly related, e.g. $q= exp (\gamma h)$ then one
has an additional dimensional parameter in the theory and the Poisson brackets
in the quasiclassical limit are highly nontrivial, e.g. for coordinates
\cite{10}

$$
\{ K_{1},K_{2} \}= \gamma ([r_{12}, K_{1} K_{2} {\cal P}_{12}] +
[K_{1} \check{r}_{12} K_{1},{\cal P}_{12}]).
$$

\noindent
In this case even the Poincar\'e group would be dynamical because its
parameters would have also nontrivial Poisson brackets (a Lie-Poisson group
\cite{3,19}). The straightforward application of the Dirac theory of the
constrained
systems results in non-autonomous equations, though with conserved momentum.
This gives rise to additional questions of interpretation if one would like to
preserve the usual mathematical structure of a physical theory \cite{20}.

{\bf Acknowledgements}: The author would like to thank J. A. de Azc\'arraga for
valuable discussions. He also wishes to thank the Spanish Ministry of
Education and Science for supporting his stay in Valencia University. This
paper is partially supported by a CICYT research grant.


\newpage

\begin{thebibliography}{99}
\itemsep 0pt
\bibitem{1} Faddeev L.D., Sklyanin E.K. and Takhtajan L.A. Teor. Matem. Fiz.
{\bf 40}, 194 (1979)
\bibitem{2} Faddeev L.D. Sov. Sci. Rev.; Sec. C: Math. Phys. {\bf 1}, 107
(1980)\\
Kulish P.P. and Sklyanin E.K. Lect. Notes Phys. {\bf 151},61 (1981)\\
Izergin A.G. and Korepin V.E. Fisika EChAYa (JINR, Dubna) {\bf 13}, 501
(1982)\\
Smirnov F.A. {\it Form factors in completely integrable models}, WS (1992)
\bibitem{3} Drinfeld V.G. Proc. ICM-86, Berkeley, {\bf 1}, 798 (1987)
\bibitem{4} Jimbo M. Lett. Math. Phys. {\bf 10}, 63 (1985)
\bibitem{5} Faddeev L.D., Reshetikhin N. Yu. and Takhtajan L.A. Algebra and
analysis {\bf 1}, 178 (1989)
\bibitem{6} Zumino B. In: {\it Mathematical Physics X} (ed. Schmudgen K.) p.
20,
Springer-Verlag, 1992
\bibitem{7} Carow-Watamura U., Schlieker M. Scholl M. and Watamura S. Z. Phys.
{\bf C48}, 159 (1990); Int. J. Mod. Phys. {\bf A6}, 3081 (1991);\\
Schmidke W., Wess J. and Zumino B. Z. Phys. {\bf C52}, 471 (1991)
\bibitem{8} Ogievetsky O., Schmidke W., Wess J. and Zumino B. Commun. Math.
Phys. {\bf 150}, 495 (1992)
\bibitem{9} de Azc\'arraga J.A., Kulish P.P. and R\'odenas F. Preprint
FTUV 93-36, Valencia (1993)
\bibitem{10} Kulish P.P. and Sklyanin E.K. J. Phys. A Math.
Gen. {\bf 25}, 5963 (1992)
\bibitem{11} Kulish P.P. and Sasaki R. Progr. Theor. Phys.
{\bf 89}, 741 (1993)
\bibitem{12} Alekseev A., Faddeev L.D. and Semenov-Tian-Shansky M.A. In:
{\it Quantum Groups}, Lect. Notes Math. {\bf 1510}, 148 (1992)
\bibitem{13} Majid S. In: {\it Quantum Groups}, Lect. Notes Math. {\bf 1510},
79 (1992);
J. Math. Phys. {\bf 32}, 3246 (1991); ibid. {\bf 34}, 2045 (1993);\\
Meyer U. Preprint DAMTP/93-45 (1993)
\bibitem{14} Kulish P.P. Teor. Matem. Fiz. {\bf 86}, 157 (1991); ibid. {\bf
94}, 193 (1993)
\bibitem{15} Rideau G. Lett. Math. Phys. {\bf 24}, 147 (1992)
\bibitem{16} Podles P. Lett. Math. Phys. {\bf 14}, 193 (1987)
\bibitem{17} Noumi M. and Mimachi K. Duke Math. Jour. {\bf 63}, 65 (1991)
\bibitem{18} Pillin M., Schmidke W. and Wess J. Nucl. Phys. {\bf B403}, 223
(1993)
\bibitem{19} Semenov-Tian-Shansky M.A. Publ. RIMS {\bf 21}, 1237 (1985)
\bibitem{20} Faddeev L.D. and Yakubowski O.A. {\it Lectures on quantum
mechanics}, LGU, Leningrad, 1975
\end{thebibliography}
\end{document}